\documentclass{appolb}
\usepackage[utf8]{inputenc}
\usepackage{epsf}
\usepackage{latexsym,amssymb,euscript}
\usepackage[dvips]{graphicx}
\usepackage[numbers,sort&compress]{natbib}
\usepackage{amsmath}
\usepackage{nicefrac}
\usepackage{slashed}
\usepackage{booktabs}
\usepackage{hyperref}
\usepackage{braket}
\usepackage{chngcntr}
\usepackage{bbold}
\usepackage{graphics}
\usepackage{graphicx}
\usepackage{mciteplus}
\usepackage{caption}
\usepackage{subcaption}
\usepackage{pdfpages}
\usepackage[titletoc]{appendix}
\graphicspath{{./figures/}}
\hypersetup{
 linktocpage = true,
 urlcolor = black,
 colorlinks = true,
 linkcolor = urlblue,
 anchorcolor = urlblue,
 citecolor = urlblue,
 pdfstartview = {XYZ null null 1.25} 
           }
\usepackage{pstricks}
\usepackage{color}
\usepackage{xcolor}
\definecolor{urlblue}{rgb}{0.2,0.4,0.7}
\definecolor{citegreen}{rgb}{0,0.4,0.2}
\definecolor{linkred}{rgb}{0.9,0.2,0.1}
\usepackage{float}
\definecolor{orcidlogocol}{HTML}{A6CE39}

\newcommand{\LQCD}{\Lambda_{\rm QCD}}

\newcommand{\tarr}{
\begin{array}}
\newcommand{\earr}{\end{array}}

\begin{document}
\title{The Higgs impact factor at next-to-leading order
\thanks{Presented at “Diffraction and Low-$x$ 2022”, Corigliano Calabro (Italy), September
24-30, 2022.}%
}
\author{Michael Fucilla, 
\address{Dipartimento di Fisica, Università della Calabria, \\ I-87036 Arcavacata di Rende, Cosenza, Italy \\ 
Istituto Nazionale di Fisica Nucleare, Gruppo collegato di Cosenza, \\ I-87036 Arcavacata di Rende,
Cosenza, Italy \\ and 
Université Paris-Saclay, CNRS/IN2P3, IJCLab, 91405, Orsay, France }
}

\maketitle
\begin{abstract}
We compute at next-to-leading order level the impact factor for the production of a forward Higgs boson from a colliding proton. 
Combined with other forward impact factors, it can be used to describe, at next-to-leading logarithmic accuracy, processes in which two objects featuring large separation in rapidity are detected at the Large Hadron Collider (LHC). As well, combined with a proper definition of the \textit{unintegrated gluon distribution} (UGD), it can be used to compute small-$x$ corrections to the forward Higgs production.     
\end{abstract}
  
\section{Introduction}
The study of high-energy reactions falling in the so-called semi-hard sector, where the Regge-Gribov scale hierarchy, $s \gg Q^2 \gg \LQCD^2$ ($s$ is the squared center-of-mass energy, $Q$ the hard scale
given by the process kinematics and $\LQCD$ the QCD mass scale) holds,
represents an excellent channel to deepen our knowledge of strong interactions in unexplored kinematic ranges. \\
It is well known that, in these kinematic conditions, the aforementioned hierarchy leads to the growth of logarithmic contributions of the form $\ln s/Q^2$, which have to be resumed to all orders.
The established tool for performing this resumation is the Balitsky–Fadin–Kuraev–Lipatov (BFKL) approach~\cite{Fadin:1975cb,Kuraev:1976ge,Kuraev:1977fs,Balitsky:1978ic}, which allows us to perform the resumation in the LLA (leading-logarithmic approximation), in which all terms $(\alpha_s \ln s)^n$ are resumed, and in the NLLA (next-to-leading logarithmic approximation), in which also terms $\alpha_s (\alpha_s \ln s)^n$ are included. The BFKL factorization allows us to express the cross sections as convolutions between a process-independent Green's function and two impact factors depicting
the transition from each incoming particle to the outgoing object(s) produced in its fragmentation region.
Although the Green's function is known in NLLA, to construct full NLLA observables, one must also know the impact factors up to next-to-leading order (NLO). 
Recent applications of the BFKL formalism to phenomenology can be found in~\cite{Ducloue:2013hia,Ducloue:2013bva,Caporale:2014gpa,Celiberto:2015yba,Celiberto:2016ygs,Caporale:2018qnm,Celiberto:2022gji,Celiberto:2016hae,Celiberto:2017ptm,Celiberto:2017ius,Bolognino:2018oth,Celiberto:2020rxb,Golec-Biernat:2018kem,Celiberto:2017nyx,Bolognino:2019yls,Celiberto:2021dzy,Celiberto:2021fdp,Boussarie:2017oae,Celiberto:2022keu,Celiberto:2022zdg,Caporale:2015vya,Caporale:2015int,Caporale:2016soq,Caporale:2016xku,Caporale:2016zkc,Celiberto:2020tmb,Hentschinski:2020tbi,Celiberto:2022fgx,Celiberto:2022qbh}. \\
In the following we consider the next-to-leading order impact factor for the production of a forward Higgs.

\section{Effective lagrangian and leading order impact factor}
We consider the next-to-leading order corrections to the Higgs impact factor in the infinite top-mass limit. Within this limit, the calculation can be greatly simplified by using the effective Lagrangian
\begin{equation}
\mathcal{L}_{ggH} = - \frac{g_H}{4} F_{\mu \nu}^{a} F^{\mu \nu,a} H \; ,
\label{EffLagrangia}
\end{equation}
where $H$ is the Higgs field, $F_{\mu \nu}^a = \partial_{\mu} A_{\nu}^a - \partial_{\nu} A_{\mu}^a  + g f^{abc} A_{\mu}^b A_{\nu}^c$ is the field strength tensor and
\begin{equation}
g_H = \frac{\alpha_s}{3 \pi v} \left( 1 + \frac{11}{4} \frac{\alpha_s}{\pi} \right) + {\cal O} (\alpha_s^3) \; .
\label{gH}
\end{equation}
At the leading order, in $D=4+2 \epsilon$ and at purely partonic level, the impact factor (differential in the Higgs kinematical variables) reads
\begin{equation}
    \frac{d\Phi_{gg}^{\{ H \}(0)} (\vec{q} \; )}{d z_H d^2 \vec{p}_H} = \frac{g_H^2}{8 (1-\epsilon) \sqrt{N^2-1}} \vec{q}^{\; 2} \delta (1-z_H) \delta^{(2)} (\vec{q}-\vec{p}_H) \; ,
    \label{LOHiggsImp2}
\end{equation}
where $z_H$ is the longitudinal fraction of the incoming gluon momenta carried by the Higgs and $q$ ($p_H$) is the Reggeon (Higgs) transverse momenta. \\

It is established that impact factors of colorless particles are infrared safe~\cite{Fadin:1999qc}, while impact factors initiated by partons suffer from infrared singularities (IRs). Since the Higgs impact factor is initiated by partons, we must introduce a factorization ansatz. 
We will write the proton initiated impact factor as a convolution between the gluon-initiated impact factor and the gluon PDF (gPDF), \emph{i.e.}
\begin{equation}
    \frac{d \Phi_{PP}^{ \{ H \}(0)} (x_H, \vec{p}_H, \vec{q})}{d x_H d^2 \vec{p}_H} = \int_{x_H}^1 \frac{d z_H}{z_H} f_g \left( \frac{x_H}{z_H} \right) \frac{d \Phi_{gg}^{ \{H \}(0)} (z_H, \vec{p}_H, \vec{q})}{d z_H d^2 \vec{p}_H}  \; ,
    \label{Factorization}
\end{equation}
where $x_H$ is the longitudinal fraction of the incoming proton momenta carried by the Higgs.
\section{Real corrections}
\begin{figure}
\begin{center}
  \includegraphics[scale=0.30]{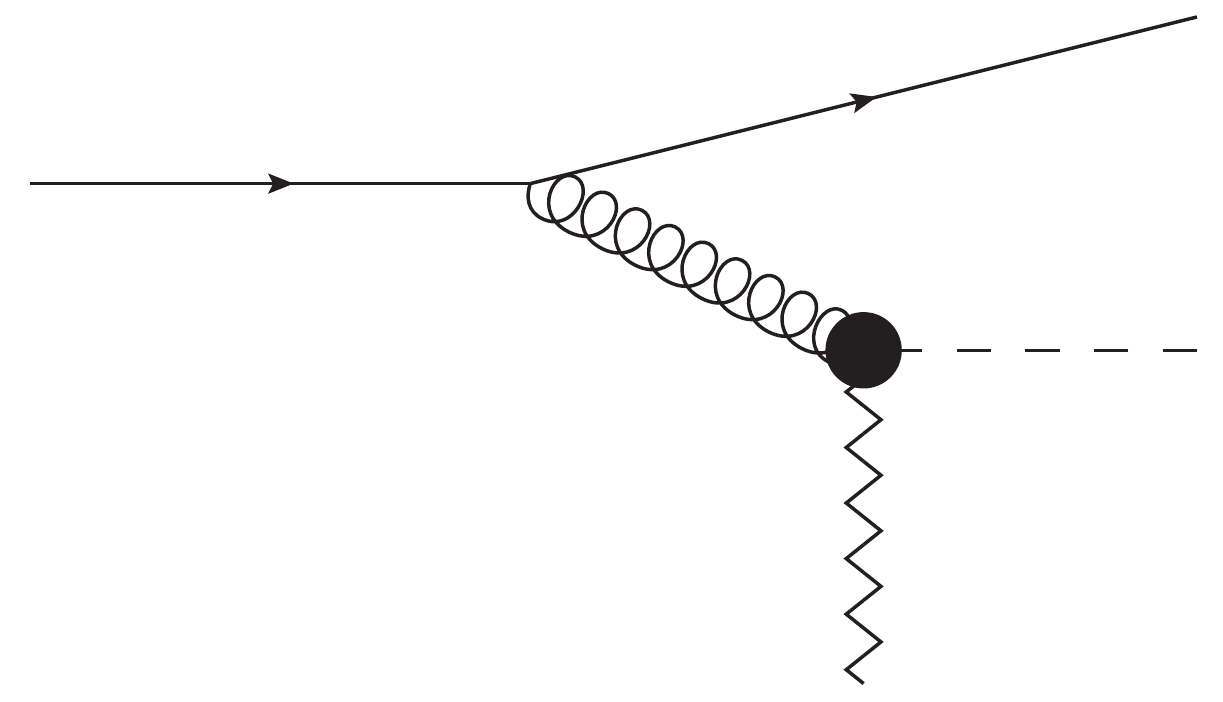} \hspace{0.5 cm}
  \includegraphics[scale=0.30]{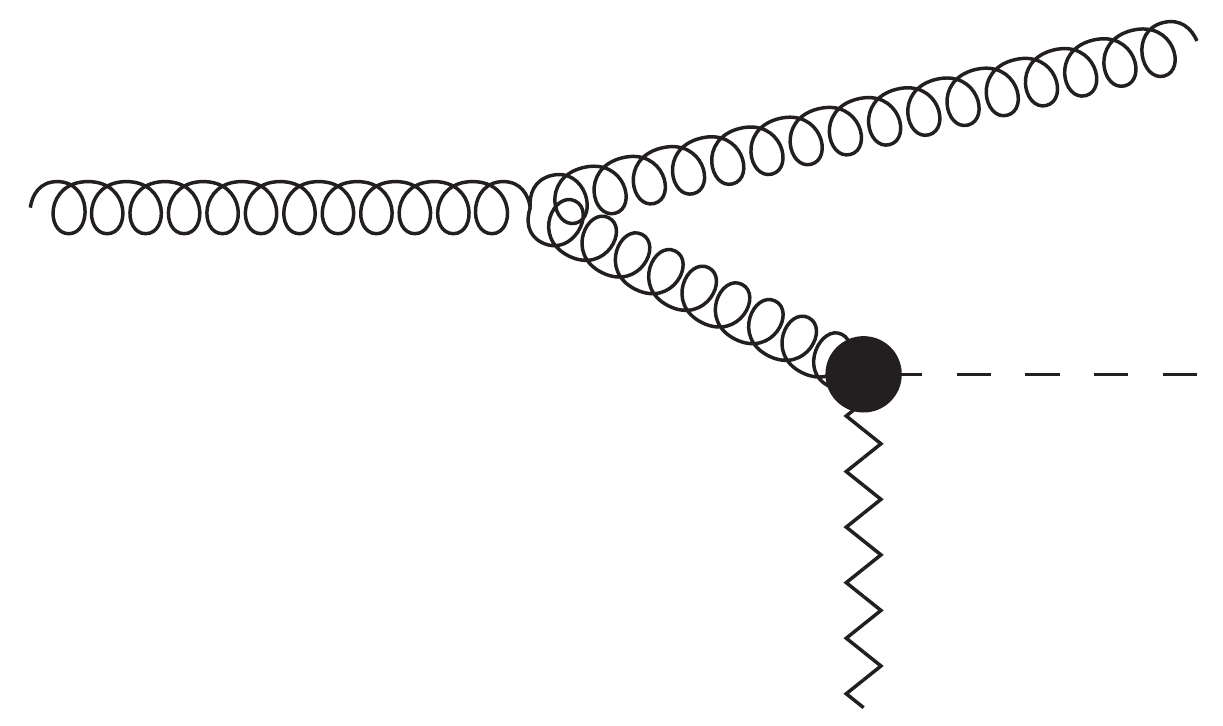} 
\end{center} 
\caption{Two diagrams contributing to real corrections.}
\label{RealCorr}
\end{figure}

At next-to-leading order the Higgs can be produced both from a quark or from a gluon (see Fig.\ref{RealCorr}). In the quark case, the correction to the impact factor is
\begin{equation}
    \frac{d \Phi_{q q}^{\{H q \}} (\vec{q} \; ) }{d z_H d^2 \vec{p}_H} = \frac{\sqrt{N^2-1}}{16 N (2 \pi)^{D-1}} \frac{g^2 g_H^2}{(\vec{r}^{\; 2})^2} \left[ \frac{4 (1-z_H) \left( \vec{r} \cdot \vec{q} \; \right)^2 + z_H^2 \vec{q}^{\; 2} \vec{r}^{\; 2}}{z_H} \right] \; ,
    \label{QuarkConImpacFin}
\end{equation}
where $\vec{r} = \vec{q}-\vec{p}_H$. As expected, we have a collinear divergences when $\vec{r}$ (transverse momenta of the outgoing quark) goes to zero. \\

In the gluon case, the correction is
\begin{equation*}
    \frac{d \Phi_{g g}^{\{H g \}} (\vec{q} \; )}{d z_H d^{2} p_H} = \frac{g^2 g_H^2 C_A}{8 (2 \pi)^{D-1}(1-\epsilon) \sqrt{N^2-1}} \frac{2 \vec{q}^{\; 2}}{\vec{r}^{\; 2}}
\end{equation*}
\begin{equation}
      \times \left[ \frac{z_H}{1-z_H} + z_H (1-z_H) + 2 (1-\epsilon) \frac{(1-z_H)}{z_H} \frac{(\vec{q} \cdot \vec{r})^2}{\vec{q}^{\; 2} \vec{r}^{\; 2}} \right]  \theta \left( s_{\Lambda} - s_{gR} \right) + \rm{finite} \; ,
     \label{GluonImp}
\end{equation}
where
\begin{equation}
    s_{gR} = \frac{(1-z_H) m_H^2 + \vec{\Delta}^2}{z_H (1-z_H)} \; , \hspace{0.5 cm} {\rm{with}} \hspace{0.5 cm} \vec{\Delta} = \vec{p}_H - z_H \vec{q} \; \; .
\end{equation}
In Eq.(\ref{GluonImp}), we have reported only the part of the impact factor which contains divergences. In this case, we have the rapidity regulator $s_{\Lambda}$ in the argument of a Heaviside theta function that prevents $z_H$ reaching the value of one\footnote{Please note that when $z_H$ reach one, the longitudinal fraction of the outgoing gluon reaches zero.} when we convolve this contribution with the gPDF as in Eq.(\ref{Factorization}). \\
In Eq.(\ref{GluonImp}), we face three types of divergence: 1) \textit{Rapidity} for $z_H \rightarrow 1$; 2) \textit{Soft} for $z_H \rightarrow 1 $ and $ \vec{r} \rightarrow 0$; 3) \textit{Collinear} for $ \vec{r} \rightarrow 0 $. \\
This result agrees with an independent calculation in the Lipatov effective action framework \cite{Hentschinski:2012kr}.

\section{Virtual corrections}
\begin{figure}
  \begin{center}
  \includegraphics[scale=0.35]{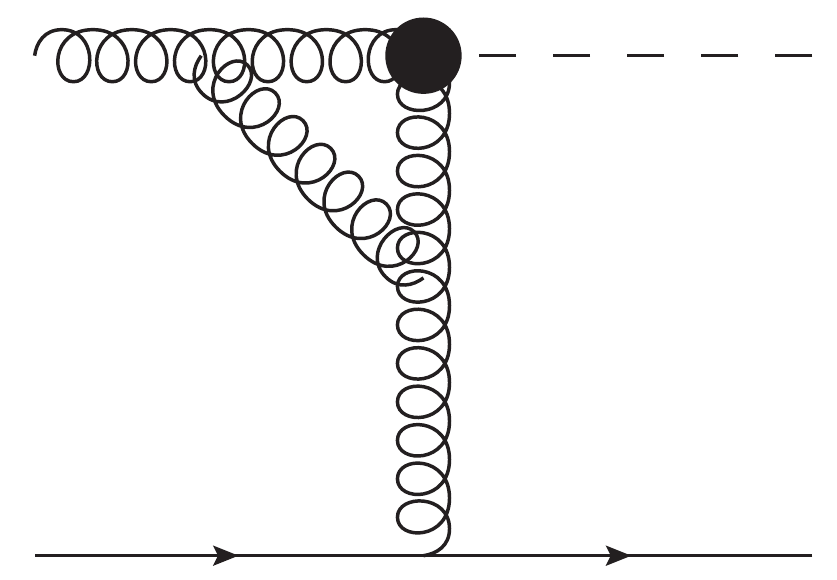} \hspace{1 cm}
  \includegraphics[scale=0.35]{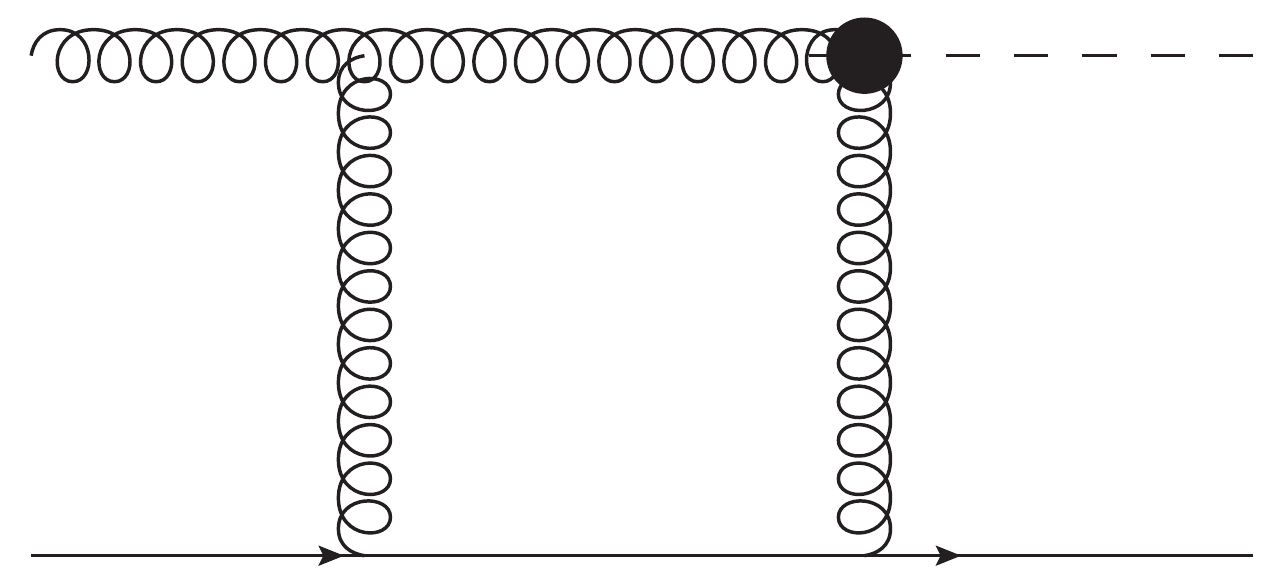} 
  \end{center}
  \caption{Two diagrams contributing to virtual corrections.}
  \label{VirtualCorr}
\end{figure}

The virtual corrections are extracted by computing the high-energy limit of a test amplitude in which the gluon quantum number are exchanged in the $t$-channel and by comparing it with the form expected from a Reggeization ansatz. In our case we use the diffusion of a gluon off a quark to produce a Higgs plus a quark. Two Feynman diagrams contributing to the process are shown in Fig.(\ref{VirtualCorr}). The virtual correction to the impact factor reads  
\begin{equation*}
\frac{d \Phi_{gg}^{\{ H \}(1)}}{d z_H d^2 \vec{p}_H} = \frac{d \Phi_{gg}^{\{ H \}(0)}}{d z_H d^2 \vec{p}_H} \; \frac{\bar{\alpha}_s}{2 \pi} \left( \frac{\vec{q}^{\; 2}}{\mu^2}  \right)^{- \epsilon}  \left[  - \frac{C_A}{\epsilon^2} + \frac{11 C_A - 2n_f}{6 \epsilon} \right.
\end{equation*}
\begin{equation}
      \left. - \frac{C_A}{\epsilon} \ln \left( \frac{\vec{q}^{\; 2}}{s_0} \right) - \frac{5 n_f}{9} + C_A \left( 2\;\Re \left( {\rm{Li}}_2 \left( 1 + \frac{m_H^2}{\vec{q}^{\; 2}} \right)\right) + \frac{\pi^2}{3} + \frac{67}{18} \right) + 11 \right] \; .
      \label{VirtualPartIMF}
\end{equation}
Once this result is established, one must observe the cancellation of divergences. An elegant method is the projection onto the LO eigenfunctions of the BFKL kernel, \emph{i.e.}
\begin{equation}
     \int \frac{d^{2-2\epsilon} q}{\pi \sqrt{2}} (\vec{q}^{\; 2})^{i \nu - \frac{3}{2}} e^{i n \phi} d \Phi _{A A}^{\left(0\right)} (\vec{q} \;) \equiv 
     d \Phi^{(0)}_{A A}(n,\nu) \; .
\label{ProjectionDef}
\end{equation}
After the projection, also divergences in the real part appear as poles in $\epsilon$ and their cancellation can be observed. The cancellation procedure takes place as follows:
\begin{itemize}
    \item The rapidity divergence appearing in the real gluon part is removed by a suitable BFKL counterterm. 
    \item Ultraviolet singularities (UVs) are taken care of by the renormalization of the QCD coupling.
    \item Soft singularities cancel out when we combine real and virtual corrections, as guaranteed by the Kinoshita-Lee-Nauenberg (KNL) theorem.
    \item The residual IRs, which are of collinear nature, are removed by renormalizing the gPDF.  
\end{itemize}
The final result can be expressed in terms of special functions and finite one-dimesional integrals of hypergeometric functions (see \cite{Celiberto:2022fgx}). \\
The virtual corrections to the impact factor in the Lipatov effective action framework has been computed in \cite{Nefedov:2019mrg}.
\section{Modification of Gribov prescription}
It is interesting to note that, the use of an effective Lagrangian containing an operator of dimension 5, requires a modification of the technique introduced in renormalizable theories for extracting the high-energy behaviour of amplitudes. In Feynman gauge, the replacement 
\begin{equation}
g^{\mu \nu} = g_{\perp\perp}^{\mu \nu} +
2 \frac{k_1^{\mu} k_2^{\nu} + k_2^{\mu} k_1^{\nu}}{s} 
\;\;\to\;\;
2 \frac{k_2^{\mu} k_1^{\nu}}{s} \; ,
\label{Gribov}
\end{equation}
known as \textit{Gribov trick}, cannot be safely applied to any $t$-channel gluon. \\
In particular, for diagrams where a single gluon connects the two rapidity regions, the trick can be safely applied, while for diagrams with two gluons exchanged in the $t$-channel, the transverse part of the metric tensor should be kept in any gluon propagator directly connected to the effective vertex. For those propagators, one should use the modified Gribov prescription, 
\begin{equation}
g^{\mu \nu} = g_{\perp\perp}^{\mu \nu} +
2 \frac{k_1^{\mu} k_2^{\nu} + k_2^{\mu} k_1^{\nu}}{s} 
\;\;\to\;\;
g_{\perp\perp}^{\mu \nu} + 2 \frac{k_2^{\mu} k_1^{\nu}}{s} \; .
\label{ModifGribov}
\end{equation}

\section{Summary and outlook}
In this work, we have presented the next-to-leading order corrections to the forward Higgs boson impact factor in the infinite top-mass limit. The current result can be directly applied to the investigation of new processes at the LHC. The immediate application of this work on the phenomenological side is to consider the Higgs plus jet production in the high-energy limit within NLLA. More formal perspectives are instead the inclusion of finite top mass corrections and the calculation of the vertex for the production of the Higgs in the central region of rapidity.

\section*{Ackowledgements}
The author is grateful to his colleagues Francesco G. Celiberto, Dmitry Yu. Ivanov, Mohammed M.A. Mohammed, Alessandro Papa, with whom this work was developed in \cite{Celiberto:2022fgx} and also to Maxim Nefedov for many useful discussion and for continuing interest.

\bibliographystyle{apsrev}
\bibliography{bibliography}

\begin{thebibliography}{36}
\expandafter\ifx\csname natexlab\endcsname\relax\def\natexlab#1{#1}\fi
\expandafter\ifx\csname bibnamefont\endcsname\relax
  \def\bibnamefont#1{#1}\fi
\expandafter\ifx\csname bibfnamefont\endcsname\relax
  \def\bibfnamefont#1{#1}\fi
\expandafter\ifx\csname citenamefont\endcsname\relax
  \def\citenamefont#1{#1}\fi
\expandafter\ifx\csname url\endcsname\relax
  \def\url#1{\texttt{#1}}\fi
\expandafter\ifx\csname urlprefix\endcsname\relax\def\urlprefix{URL }\fi
\providecommand{\bibinfo}[2]{#2}
\providecommand{\eprint}[2][]{\url{#2}}

\bibitem[{\citenamefont{Fadin et~al.}(1975)}]{Fadin:1975cb}
\bibinfo{author}{\bibfnamefont{V.~S.} \bibnamefont{Fadin}}
  \bibnamefont{et~al.}, \bibinfo{journal}{Phys. Lett. B}
  \textbf{\bibinfo{volume}{60}}, \bibinfo{pages}{50} (\bibinfo{year}{1975}).

\bibitem[{\citenamefont{Kuraev et~al.}(1976)}]{Kuraev:1976ge}
\bibinfo{author}{\bibfnamefont{E.~A.} \bibnamefont{Kuraev}}
  \bibnamefont{et~al.}, \bibinfo{journal}{Sov. Phys. JETP}
  \textbf{\bibinfo{volume}{44}}, \bibinfo{pages}{443} (\bibinfo{year}{1976}).

\bibitem[{\citenamefont{Kuraev et~al.}(1977)}]{Kuraev:1977fs}
\bibinfo{author}{\bibfnamefont{E.}~\bibnamefont{Kuraev}} \bibnamefont{et~al.},
  \bibinfo{journal}{Sov.\ Phys.\ JETP} \textbf{\bibinfo{volume}{45}},
  \bibinfo{pages}{199} (\bibinfo{year}{1977}).

\bibitem[{\citenamefont{Balitsky and Lipatov}(1978)}]{Balitsky:1978ic}
\bibinfo{author}{\bibfnamefont{I.}~\bibnamefont{Balitsky}} \bibnamefont{and}
  \bibinfo{author}{\bibfnamefont{L.}~\bibnamefont{Lipatov}},
  \bibinfo{journal}{Sov.\ J.\ Nucl.\ Phys.} \textbf{\bibinfo{volume}{28}},
  \bibinfo{pages}{822} (\bibinfo{year}{1978}).

\bibitem[{\citenamefont{Duclou\'e et~al.}(2013)}]{Ducloue:2013hia}
\bibinfo{author}{\bibfnamefont{B.}~\bibnamefont{Duclou\'e}}
  \bibnamefont{et~al.}, \bibinfo{journal}{JHEP} \textbf{\bibinfo{volume}{05}},
  \bibinfo{pages}{096} (\bibinfo{year}{2013}), \eprint{1302.7012}.

\bibitem[{\citenamefont{Duclou\'e et~al.}(2014)}]{Ducloue:2013bva}
\bibinfo{author}{\bibfnamefont{B.}~\bibnamefont{Duclou\'e}}
  \bibnamefont{et~al.}, \bibinfo{journal}{Phys. Rev. Lett.}
  \textbf{\bibinfo{volume}{112}}, \bibinfo{pages}{082003}
  (\bibinfo{year}{2014}), \eprint{1309.3229}.

\bibitem[{\citenamefont{Caporale et~al.}(2014)}]{Caporale:2014gpa}
\bibinfo{author}{\bibfnamefont{F.}~\bibnamefont{Caporale}}
  \bibnamefont{et~al.}, \bibinfo{journal}{Eur. Phys. J. C}
  \textbf{\bibinfo{volume}{74}}, \bibinfo{pages}{3084} (\bibinfo{year}{2014}),
  \eprint{1407.8431}.

\bibitem[{\citenamefont{Celiberto et~al.}(2015)}]{Celiberto:2015yba}
\bibinfo{author}{\bibfnamefont{F.~G.} \bibnamefont{Celiberto}}
  \bibnamefont{et~al.}, \bibinfo{journal}{Eur. Phys. J. C}
  \textbf{\bibinfo{volume}{75}}, \bibinfo{pages}{292} (\bibinfo{year}{2015}),
  \eprint{1504.08233}.

\bibitem[{\citenamefont{Celiberto
  et~al.}(2016{\natexlab{a}})}]{Celiberto:2016ygs}
\bibinfo{author}{\bibfnamefont{F.~G.} \bibnamefont{Celiberto}}
  \bibnamefont{et~al.}, \bibinfo{journal}{Eur. Phys. J. C}
  \textbf{\bibinfo{volume}{76}}, \bibinfo{pages}{224}
  (\bibinfo{year}{2016}{\natexlab{a}}), \eprint{1601.07847}.

\bibitem[{\citenamefont{Caporale et~al.}(2018)}]{Caporale:2018qnm}
\bibinfo{author}{\bibfnamefont{F.}~\bibnamefont{Caporale}}
  \bibnamefont{et~al.}, \bibinfo{journal}{Nucl. Phys. B}
  \textbf{\bibinfo{volume}{935}}, \bibinfo{pages}{412} (\bibinfo{year}{2018}),
  \eprint{1806.06309}.

\bibitem[{\citenamefont{Celiberto
  et~al.}(2022{\natexlab{a}})}]{Celiberto:2022gji}
\bibinfo{author}{\bibfnamefont{F.~G.} \bibnamefont{Celiberto}}
  \bibnamefont{et~al.}, \bibinfo{journal}{Phys. Rev. D, in press}
  (\bibinfo{year}{2022}{\natexlab{a}}), \eprint{2207.05015}.

\bibitem[{\citenamefont{Celiberto
  et~al.}(2016{\natexlab{b}})}]{Celiberto:2016hae}
\bibinfo{author}{\bibfnamefont{F.~G.} \bibnamefont{Celiberto}}
  \bibnamefont{et~al.}, \bibinfo{journal}{Phys. Rev. D}
  \textbf{\bibinfo{volume}{94}}, \bibinfo{pages}{034013}
  (\bibinfo{year}{2016}{\natexlab{b}}), \eprint{1604.08013}.

\bibitem[{\citenamefont{Celiberto et~al.}(2017)}]{Celiberto:2017ptm}
\bibinfo{author}{\bibfnamefont{F.~G.} \bibnamefont{Celiberto}}
  \bibnamefont{et~al.}, \bibinfo{journal}{Eur. Phys. J. C}
  \textbf{\bibinfo{volume}{77}}, \bibinfo{pages}{382} (\bibinfo{year}{2017}),
  \eprint{1701.05077}.

\bibitem[{\citenamefont{Celiberto}(2017)}]{Celiberto:2017ius}
\bibinfo{author}{\bibfnamefont{F.~G.} \bibnamefont{Celiberto}}, Ph.D. thesis
  (\bibinfo{year}{2017}), \eprint{1707.04315}.

\bibitem[{\citenamefont{Bolognino et~al.}(2018)}]{Bolognino:2018oth}
\bibinfo{author}{\bibfnamefont{A.~D.} \bibnamefont{Bolognino}}
  \bibnamefont{et~al.}, \bibinfo{journal}{Eur. Phys. J. C}
  \textbf{\bibinfo{volume}{78}}, \bibinfo{pages}{772} (\bibinfo{year}{2018}),
  \eprint{1808.05483}.

\bibitem[{\citenamefont{Celiberto et~al.}(2020)}]{Celiberto:2020rxb}
\bibinfo{author}{\bibfnamefont{F.~G.} \bibnamefont{Celiberto}}
  \bibnamefont{et~al.}, \bibinfo{journal}{Phys. Rev. D}
  \textbf{\bibinfo{volume}{102}}, \bibinfo{pages}{094019}
  (\bibinfo{year}{2020}), \eprint{2008.10513}.

\bibitem[{\citenamefont{Golec-Biernat et~al.}(2018)}]{Golec-Biernat:2018kem}
\bibinfo{author}{\bibfnamefont{K.}~\bibnamefont{Golec-Biernat}}
  \bibnamefont{et~al.}, \bibinfo{journal}{JHEP} \textbf{\bibinfo{volume}{12}},
  \bibinfo{pages}{091} (\bibinfo{year}{2018}), \eprint{1811.04361}.

\bibitem[{\citenamefont{Celiberto et~al.}(2018)}]{Celiberto:2017nyx}
\bibinfo{author}{\bibfnamefont{F.~G.} \bibnamefont{Celiberto}}
  \bibnamefont{et~al.}, \bibinfo{journal}{Phys. Lett. B}
  \textbf{\bibinfo{volume}{777}}, \bibinfo{pages}{141} (\bibinfo{year}{2018}),
  \eprint{1709.10032}.

\bibitem[{\citenamefont{Bolognino et~al.}(2019)}]{Bolognino:2019yls}
\bibinfo{author}{\bibfnamefont{A.~D.} \bibnamefont{Bolognino}}
  \bibnamefont{et~al.}, \bibinfo{journal}{Eur. Phys. J. C}
  \textbf{\bibinfo{volume}{79}}, \bibinfo{pages}{939} (\bibinfo{year}{2019}),
  \eprint{1909.03068}.

\bibitem[{\citenamefont{Celiberto
  et~al.}(2021{\natexlab{a}})}]{Celiberto:2021dzy}
\bibinfo{author}{\bibfnamefont{F.~G.} \bibnamefont{Celiberto}}
  \bibnamefont{et~al.}, \bibinfo{journal}{Eur. Phys. J. C}
  \textbf{\bibinfo{volume}{81}}, \bibinfo{pages}{780}
  (\bibinfo{year}{2021}{\natexlab{a}}), \eprint{2105.06432}.

\bibitem[{\citenamefont{Celiberto
  et~al.}(2021{\natexlab{b}})}]{Celiberto:2021fdp}
\bibinfo{author}{\bibfnamefont{F.~G.} \bibnamefont{Celiberto}}
  \bibnamefont{et~al.}, \bibinfo{journal}{Phys. Rev. D}
  \textbf{\bibinfo{volume}{104}}, \bibinfo{pages}{114007}
  (\bibinfo{year}{2021}{\natexlab{b}}), \eprint{2109.11875}.

\bibitem[{\citenamefont{Boussarie et~al.}(2018)}]{Boussarie:2017oae}
\bibinfo{author}{\bibfnamefont{R.}~\bibnamefont{Boussarie}}
  \bibnamefont{et~al.}, \bibinfo{journal}{Phys. Rev. D}
  \textbf{\bibinfo{volume}{97}}, \bibinfo{pages}{014008}
  (\bibinfo{year}{2018}), \eprint{1709.01380}.

\bibitem[{\citenamefont{Celiberto}(2022)}]{Celiberto:2022keu}
\bibinfo{author}{\bibfnamefont{F.~G.} \bibnamefont{Celiberto}},
  \bibinfo{journal}{Phys. Lett. B} \textbf{\bibinfo{volume}{835}},
  \bibinfo{pages}{137554} (\bibinfo{year}{2022}), \eprint{2206.09413}.

\bibitem[{\citenamefont{Celiberto
  et~al.}(2022{\natexlab{b}})}]{Celiberto:2022zdg}
\bibinfo{author}{\bibfnamefont{F.~G.} \bibnamefont{Celiberto}}
  \bibnamefont{et~al.}, \bibinfo{journal}{Phys. Rev. D}
  \textbf{\bibinfo{volume}{105}}, \bibinfo{pages}{114056}
  (\bibinfo{year}{2022}{\natexlab{b}}), \eprint{2205.13429}.

\bibitem[{\citenamefont{Caporale
  et~al.}(2016{\natexlab{a}})}]{Caporale:2015vya}
\bibinfo{author}{\bibfnamefont{F.}~\bibnamefont{Caporale}}
  \bibnamefont{et~al.}, \bibinfo{journal}{Phys. Rev. Lett.}
  \textbf{\bibinfo{volume}{116}}, \bibinfo{pages}{012001}
  (\bibinfo{year}{2016}{\natexlab{a}}), \eprint{1508.07711}.

\bibitem[{\citenamefont{Caporale
  et~al.}(2016{\natexlab{b}})}]{Caporale:2015int}
\bibinfo{author}{\bibfnamefont{F.}~\bibnamefont{Caporale}}
  \bibnamefont{et~al.}, \bibinfo{journal}{Eur. Phys. J. C}
  \textbf{\bibinfo{volume}{76}}, \bibinfo{pages}{165}
  (\bibinfo{year}{2016}{\natexlab{b}}), \eprint{1512.03364}.

\bibitem[{\citenamefont{Caporale
  et~al.}(2016{\natexlab{c}})}]{Caporale:2016soq}
\bibinfo{author}{\bibfnamefont{F.}~\bibnamefont{Caporale}}
  \bibnamefont{et~al.}, \bibinfo{journal}{Nucl. Phys. B}
  \textbf{\bibinfo{volume}{910}}, \bibinfo{pages}{374}
  (\bibinfo{year}{2016}{\natexlab{c}}), \eprint{1603.07785}.

\bibitem[{\citenamefont{Caporale
  et~al.}(2017{\natexlab{a}})}]{Caporale:2016xku}
\bibinfo{author}{\bibfnamefont{F.}~\bibnamefont{Caporale}}
  \bibnamefont{et~al.}, \bibinfo{journal}{Eur. Phys. J. C}
  \textbf{\bibinfo{volume}{77}}, \bibinfo{pages}{5}
  (\bibinfo{year}{2017}{\natexlab{a}}), \eprint{1606.00574}.

\bibitem[{\citenamefont{Caporale
  et~al.}(2017{\natexlab{b}})}]{Caporale:2016zkc}
\bibinfo{author}{\bibfnamefont{F.}~\bibnamefont{Caporale}}
  \bibnamefont{et~al.}, \bibinfo{journal}{Phys. Rev. D}
  \textbf{\bibinfo{volume}{95}}, \bibinfo{pages}{074007}
  (\bibinfo{year}{2017}{\natexlab{b}}), \eprint{1612.05428}.

\bibitem[{\citenamefont{Celiberto
  et~al.}(2021{\natexlab{c}})}]{Celiberto:2020tmb}
\bibinfo{author}{\bibfnamefont{F.~G.} \bibnamefont{Celiberto}}
  \bibnamefont{et~al.}, \bibinfo{journal}{Eur. Phys. J. C}
  \textbf{\bibinfo{volume}{81}}, \bibinfo{pages}{293}
  (\bibinfo{year}{2021}{\natexlab{c}}), \eprint{2008.00501}.

\bibitem[{\citenamefont{Hentschinski et~al.}(2021)}]{Hentschinski:2020tbi}
\bibinfo{author}{\bibfnamefont{M.}~\bibnamefont{Hentschinski}}
  \bibnamefont{et~al.}, \bibinfo{journal}{Eur. Phys. J. C}
  \textbf{\bibinfo{volume}{81}}, \bibinfo{pages}{112} (\bibinfo{year}{2021}),
  \eprint{2011.03193}.

\bibitem[{\citenamefont{Celiberto
  et~al.}(2022{\natexlab{c}})}]{Celiberto:2022fgx}
\bibinfo{author}{\bibfnamefont{F.~G.} \bibnamefont{Celiberto}}
  \bibnamefont{et~al.}, \bibinfo{journal}{JHEP} \textbf{\bibinfo{volume}{08}},
  \bibinfo{pages}{092} (\bibinfo{year}{2022}{\natexlab{c}}),
  \eprint{2205.02681}.

\bibitem[{\citenamefont{Celiberto
  et~al.}(2022{\natexlab{d}})\citenamefont{Celiberto, Fucilla, and
  Papa}}]{Celiberto:2022qbh}
\bibinfo{author}{\bibfnamefont{F.~G.} \bibnamefont{Celiberto}},
  \bibinfo{author}{\bibfnamefont{M.}~\bibnamefont{Fucilla}}, \bibnamefont{and}
  \bibinfo{author}{\bibfnamefont{A.}~\bibnamefont{Papa}}
  (\bibinfo{year}{2022}{\natexlab{d}}), \eprint{2209.01372}.

\bibitem[{\citenamefont{Fadin and Martin}(1999)}]{Fadin:1999qc}
\bibinfo{author}{\bibfnamefont{V.~S.} \bibnamefont{Fadin}} \bibnamefont{and}
  \bibinfo{author}{\bibfnamefont{A.~D.} \bibnamefont{Martin}},
  \bibinfo{journal}{Phys. Rev. D} \textbf{\bibinfo{volume}{60}},
  \bibinfo{pages}{114008} (\bibinfo{year}{1999}), \eprint{hep-ph/9904505}.

\bibitem[{\citenamefont{Hentschinski and
  others}(2013)\citenamefont{Hentschinski et~al.}}]{Hentschinski:2012kr}
\bibinfo{author}{\bibfnamefont{M.}~\bibnamefont{Hentschinski}}
  \bibnamefont{et~al.}, \bibinfo{journal}{Phys. Rev. Lett.}
  \textbf{\bibinfo{volume}{110}}, \bibinfo{pages}{041601}
  (\bibinfo{year}{2013}), \eprint{1209.1353}.

\bibitem[{\citenamefont{Nefedov}(2019)}]{Nefedov:2019mrg}
\bibinfo{author}{\bibfnamefont{M.~A.} \bibnamefont{Nefedov}},
  \bibinfo{journal}{Nucl. Phys. B} \textbf{\bibinfo{volume}{946}},
  \bibinfo{pages}{114715} (\bibinfo{year}{2019}), \eprint{1902.11030}.

\end{thebibliography}

\end{document}